\documentclass[aps,reprint,prl,showpacs,superscriptaddress]{revtex4-2}
\usepackage{amsmath}
\usepackage{amssymb}
\usepackage{graphicx}
\usepackage{color}
\usepackage{bm}

\newcommand{\qq}{\begin{eqnarray}}
\newcommand{\qqq}{\end{eqnarray}}

\newcommand{\p}{\partial}

\newcommand{\bfr}{\bm{r}}

\usepackage{babel}

\begin{document}

\title{Absorbing phase transitions in systems with mediated interactions}

\author{Romain Mari}
\author{Eric Bertin} 
\affiliation{Universit\'e Grenoble Alpes \& CNRS, LIPhy, 38000 Grenoble, France}

\author{Cesare Nardini} 
\affiliation{Service de Physique de l'\'Etat Condens\'e, CNRS UMR 3680, CEA-Saclay, 91191 Gif-sur-Yvette, France}

\date{\today}

\begin{abstract}
Experiments of periodically sheared colloidal suspensions or  soft amorphous solids display a transition from reversible to irreversible particle motion that, when analysed stroboscopically in time, is interpreted as an absorbing phase transition with infinitely many absorbing states. In these systems interactions mediated by hydrodynamics or elasticity are present, causing passive regions to be affected by nearby active ones. We show that mediated interactions induce a new universality class of absorbing phase transitions, distinct from Conserved Directed Percolation, and we obtain the corresponding critical exponents.
We do so with large-scale numerical simulations of a minimal model for the stroboscopic dynamics of sheared soft materials and derive the minimal field theoretical description.
\end{abstract}


\maketitle






Soft materials under cyclic shear often show an intriguing phase transition in their microscopic dynamics, termed Reversible-Irreversible Transition (RIT). 
At low enough driving amplitudes, the system reaches a reversible state where its configuration is strictly unchanged when observed stroboscopically (once per cycle), whereas at large amplitudes the stroboscopic dynamics is diffusive. 
This transition is observed in systems varying considerably in their microscopics, including non-Brownian suspensions~\cite{pine2005chaos,corte2009self}, granular materials~\cite{royer2015precisely}, microemulsions~\cite{weijs2015emergent} and soft glasses~\cite{fiocco2013oscillatory}.
While for dilute suspensions reversibility is borne from time-reversible Stokes hydrodynamics~\cite{corte2008random}, 
for jammed systems it is argued to come from a repeated sequence of plastic events~\cite{munganCyclicAnnealingIterated2019,munganNetworksHierarchiesHow2019}.

The RIT is a form of Absorbing Phase Transition (APT). 
The order parameter is the activity $A$, which measures the fraction of the system which evolves irreversibly during a cycle; reversible states correspond to $A=0$.
APTs arise in many non-equilibrium contexts, among which the spreading of infectious diseases, reaction-diffusion problems and fracture propagation~\cite{hinrichsen2000non}.
In the case of RIT, there exist infinitely many absorbing states not related by any symmetry (e.g., all configurations leading to contact-free cycles for dilute suspensions), 
and the particle number $N$ is conserved. 
It has therefore been argued to belong to the Conserved Directed Percolation (CDP) or Manna class~\cite{hinrichsen2000non,manna1991two,vespignani1998driving,rossi2000universality,corte2008random,le2015exact,ness2020absorbing}. 
This class is described at field-theoretical level~\cite{vespignani1998driving,van2002universality,rossi2000universality,le2015exact} by the local density $\rho(\bfr, t)$ (normalised such that $\int d\bfr \rho(\bfr,t) =N=V\rho_0$) and activity $A(\bfr,t)$, with dynamics
\begin{align}
\p_t \rho & = D_{\rho}\nabla^2 A \label{eq:Manna-standard-1}\\
\p_t A    & = f(A) + D_A \nabla^2 A + \sigma \sqrt{A}\eta\, ,
\label{eq:Manna-standard-2}
\end{align}
where $f(A) = f_\mathrm{CDP}(A) \equiv (-\alpha+k\rho)A - \lambda A^2$ and $\eta(\bfr, t)$ is a Gaussian noise with zero average and $\langle \eta(\bfr, t)\eta(\bfr',t')\rangle = \delta(\bfr-\bfr')\delta(t-t')$. 
Neglecting the noise term in Eq.~\eqref{eq:Manna-standard-2}, the transition to absorbing states is located at $\rho_0^c=\alpha/k$ 
where the mean activity vanishes as $\langle A \rangle \sim (\rho_0-\rho_0^c)^{\beta}$. 
The mean-field exponent $\beta_\mathrm{CDP}^\mathrm{MF}=1$ is larger than the value $\beta_\mathrm{CDP} \approx 0.64$ found in two dimensions~\cite{lubeck2004universal}.

However, while CDP has been argued to capture some realizations of RIT~\cite{corte2008random,himanagamanasaExperimentalSignaturesNonequilibrium2014}, several results challenge 
the CDP classification of the RIT. 
Some experiments report a convex behavior ($\beta>1$) close the transition, both above~\cite{knowlton_microscopic_2014} (consistent with numerical results~\cite{regev_critical_2018}) 
and below jamming~\cite{weijs2015emergent}.
Others report a first order transition in semi-dilute systems~\cite{jeanneret_geometrically_2014}, when several numerical works also  report it close to or above jamming~\cite{kawasaki2016macroscopic,nagasawa_classification_2019,parmar_strain_2019,das_unified_2020,yeh_glass_2020}.
Some of these CDP-incompatible behaviors have been argued to stem from hydrodynamic interactions~\cite{weijs2015emergent}.
More generally, a natural expectation is that due to hydrodynamic or elastic interactions, a region of local activity can impact nearby passive ones.
This mechanism is absent from both CDP field theory and minimal models implementing it~\cite{corte2008random,tjhung2015hyperuniform,ness2020absorbing}. 

In this Letter we report a new universality class for APTs with infinitely many absorbing states. It is distinct from CDP and arises when passive regions are affected by active ones. 
We do so by introducing a generalization of the minimal model studied in~\cite{tjhung2015hyperuniform} for the stroboscopic dynamics of periodically sheared suspensions. 
The effect of mediated interactions is mimicked at a mean-field level by a diffusion of passive particles proportional to the total activity. 
We characterize the RIT for our model in simulations and show that activity-induced diffusion of passive particles makes it either a convex second-order transition with $\beta>1$ 
or even a first order transition. Furthermore, by a coarse-graining of our minimal model we show that the CDP normal form~\ref{eq:Manna-standard-2} is replaced by
\begin{equation}
f(A) = f_\mathrm{CDP}(A)+f_\mathrm{p}(A) = (-\alpha+\tilde{k}\rho)A -\mu \rho A^{3/2} - \lambda A^2,
\end{equation}
where $\tilde{k}$ is a renormalized coefficient. The presence of a $A^{3/2}$ term, with $\mu>0$, is key to the new universality class. 
Such non-analytic term eludes symmetry or conservation arguments; this property is shared by other non-equilibrium phase transitions~\cite{aron2020nonanalytic} and is probably the reason for which this universality class was missed by previous works.

\begin{figure}[t]
\begin{centering}
\includegraphics[width=0.8\columnwidth]{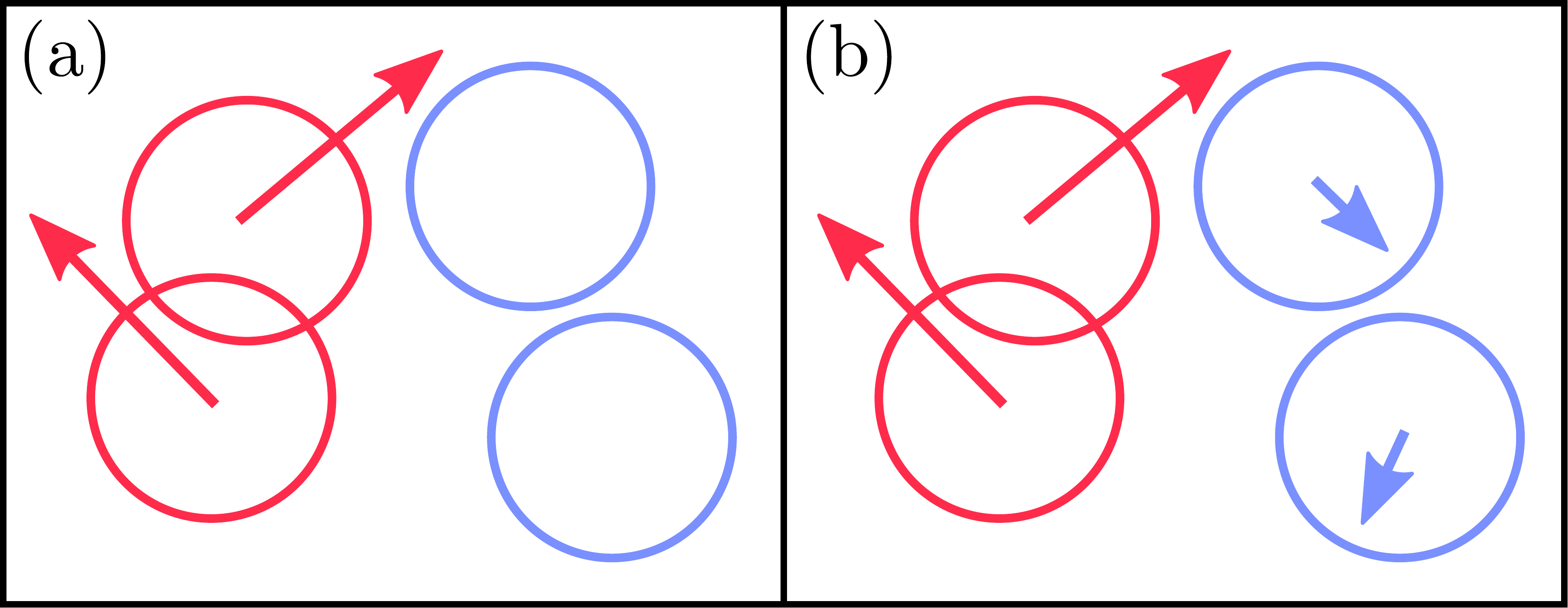}
\par\end{centering}
\caption{(a) Tjhung-Berthier model: particles in overlap (``active'', in red) at a given time step are moved with a random vector with typical length $\Delta^{\rm a}$, particles without overlap (``passive'', in blue) are kept fixed. (b) Our model: active particles follow the same dynamics than in TB model, whereas passive particles are also moved with a random vector with typical length $\Delta^{\rm p}$ function of the total activity.}
\label{fig:model-definition}
\end{figure}

\paragraph{Minimal particle model}
Our starting point is the minimal model proposed by Tjhung and Berthier (TB)~\cite{tjhung2015hyperuniform}. 
A set of $N$ disks of diameter $D$ are distributed in space (we here consider only the $2$-dimensional case); those overlapping with others are called ``active'', the others are ``passive''. 
With reference to a dilute suspension, active particles are interpreted as particles that collide during a shear cycle in the real system.
At each time step, active particles are moved by a Gaussian-distributed random displacement with standard deviation $\Delta_\mathrm{a}$. 
In the original TB model, passive particles are kept fixed [Fig.~\ref{fig:model-definition}(a)]. 
We instead assume that passive particles are randomly displaced over a distance that depends on the overall activity at time $t$, $\bar{A}(t)=N_\mathrm{a}(t)/N$ (with $N_\mathrm{a}$ the number of active particles) [Fig.~\ref{fig:model-definition}(b)].  
Indeed, in a real system a displacement $\bm{\Delta}^\mathrm{a}_i$ of an active particle $i$ induces a displacement $\bm{\Delta}^\mathrm{p}_j$ on a passive particle $j$ 
separated by $\bm{r}_{ij}$ via a tensorial propagator $\bm{G}(\bm{r}_{ij})$ (which may be long-ranged), such that $\bm{\Delta}^\mathrm{p}_j = \bm{G}(\bm{r}_{ij})\!\cdot\!\bm{\Delta}^\mathrm{a}_i$. 
We assume that the total displacement of a passive particle generated by several active particles is additive.
Further assuming that active displacements are uncorrelated, one obtains that the variance of passive displacements 
is $\langle (\bm{\Delta}_i^\mathrm{p})^2 \rangle = \Delta_\mathrm{a}^2 K \rho_0 \bar{A}$, with $K=\int \mathrm{d}\bm{r}\, \bm{G}(\bm{r})\!\! :\!\! \bm{G}(\bm{r})$, 
assuming this integral converges.
This holds even for long-ranged interactions ($G_{\alpha\beta}(\bm{r}) \sim 1/r^{\mu}$ at large distance $r$) as far as $\mu>\frac{d}{2}$, a condition satisfied by the elastic propagator ($\mu=d$) and by hydrodynamic interactions caused by force dipoles such as particle contacts. 
More generally, relaxing the assumptions of uncorrelated active motion will lead to corrections to the variance of order $\bar{A}^{2}$. 
We thus account for the effect of mediated interactions in a mean-field spirit by assuming Gaussian-distributed displacements of passive particles with a standard deviation $\Delta_\mathrm{p} =\lambda_\mathrm{p}\sqrt{\bar{A}}\left[1+c\bar{A}\right]$, where $\lambda_\mathrm{p}>0$ and $c>-1$ to ensure positivity.

\paragraph{Numerical analysis.--}
For $\lambda_\mathrm{p}=0$, our model reduces to the TB model, which is expected to belong to the CDP universality class; we confirm this expectation below. 
To focus on the potential deviations to CDP, we here restrict to $\Delta_\mathrm{a}/D=1$.
Dimensionless model parameters are then (i) the area fraction $\phi=\pi D^2 \rho_0/4$, (ii) the ratio $\lambda_\mathrm{p}/D$, and (iii) the coefficient $c$ governing the correction to the scaling $\Delta \sim \sqrt{\bar{A}}$.
We use system sizes up to $N=2^{24}$ to explore the phase diagram as close as possible to the APT.
We first focus on the case $c=0$, and briefly discuss later on the effect of $c\ne 0$.

\begin{figure}[t]
\begin{centering}
\includegraphics[width=1\columnwidth]{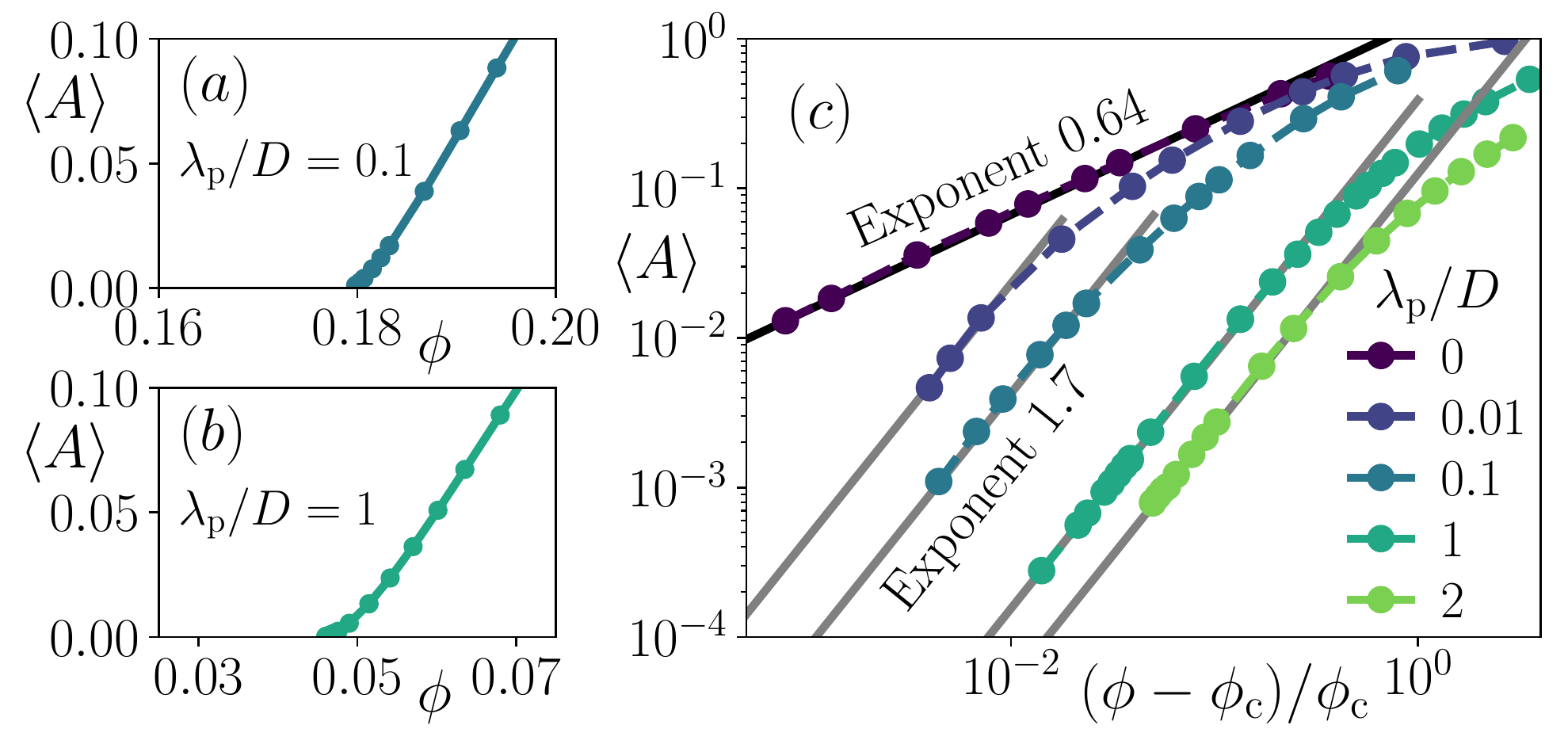}
\par\end{centering}
\caption{Left: average activity $\langle A\rangle$ versus mean area fraction $\phi$ for $c=0$, and (a) $\lambda_\mathrm{p}/D = 0.1$ or (b) $\lambda_\mathrm{p}/D = 1$. (c) $\log \langle A\rangle$ versus $\log (\phi-\phi_\mathrm{c})/\phi_\mathrm{c}$, showing the critical behavior at $c=0$, for several values of $\lambda_\mathrm{p}/D$.}
\label{fig:AvsRho}
\end{figure}

We plot in Fig.~\ref{fig:AvsRho}(a) and (b) the average activity in steady-state $\langle A \rangle$ as a function of $\phi$, 
for two different values of $\lambda_\mathrm{p}/D$, for $c=0$.
We observe an APT at a value $\phi_\mathrm{c}$ which decreases when increasing $\lambda_\mathrm{p}/D$, that is, with stronger mediated interactions.
Importantly, the convexity of $\langle A \rangle(\phi)$ observed in Fig.~\ref{fig:AvsRho}(a) and (b) indicates that 
$\beta>1$, in contrast with the CDP value $\beta_\mathrm{CDP} \approx 0.64$.
This is confirmed on a logarithmic scale showing $\langle A \rangle$ as a function of $\varepsilon \equiv (\phi-\phi_\mathrm{c})/\phi_\mathrm{c}$ in Fig.~\ref{fig:AvsRho}(c).
(We show in~\cite{supp} how we determined $\phi_\mathrm{c}$.)
For $\lambda_\mathrm{p}/D=0$ (TB model), we recover $\beta_\mathrm{CDP} \approx 0.64$.
For $\lambda_\mathrm{p}/D > 0$, a crossover is observed between a regime compatible with $\beta_\mathrm{CDP}$ (at least for small $\lambda_\mathrm{p}/D$) 
far enough from the critical point, and a new critical behavior with $\beta \approx 1.7$ in an interval close to the critical point, which widens upon increase of $\lambda_\mathrm{p}/D$.
It thus appears clearly that a tiny motion of passive particles mimicking the effect of mediated interactions, modifies the universality class of the APT.
Quite importantly, the value $\beta \approx 1.7$ is much larger than the mean-field value $\beta_\mathrm{CDP}^\mathrm{MF}=1$ showing that, 
although we have included mediated interactions in a mean-field manner, the model is not following mean-field directed percolation, 
which rather happens when $\Delta_\mathrm{a}/D \to \infty$~\cite{lubeck2004universal}.

\begin{figure}[t]
\begin{centering}
\includegraphics[width=\columnwidth]{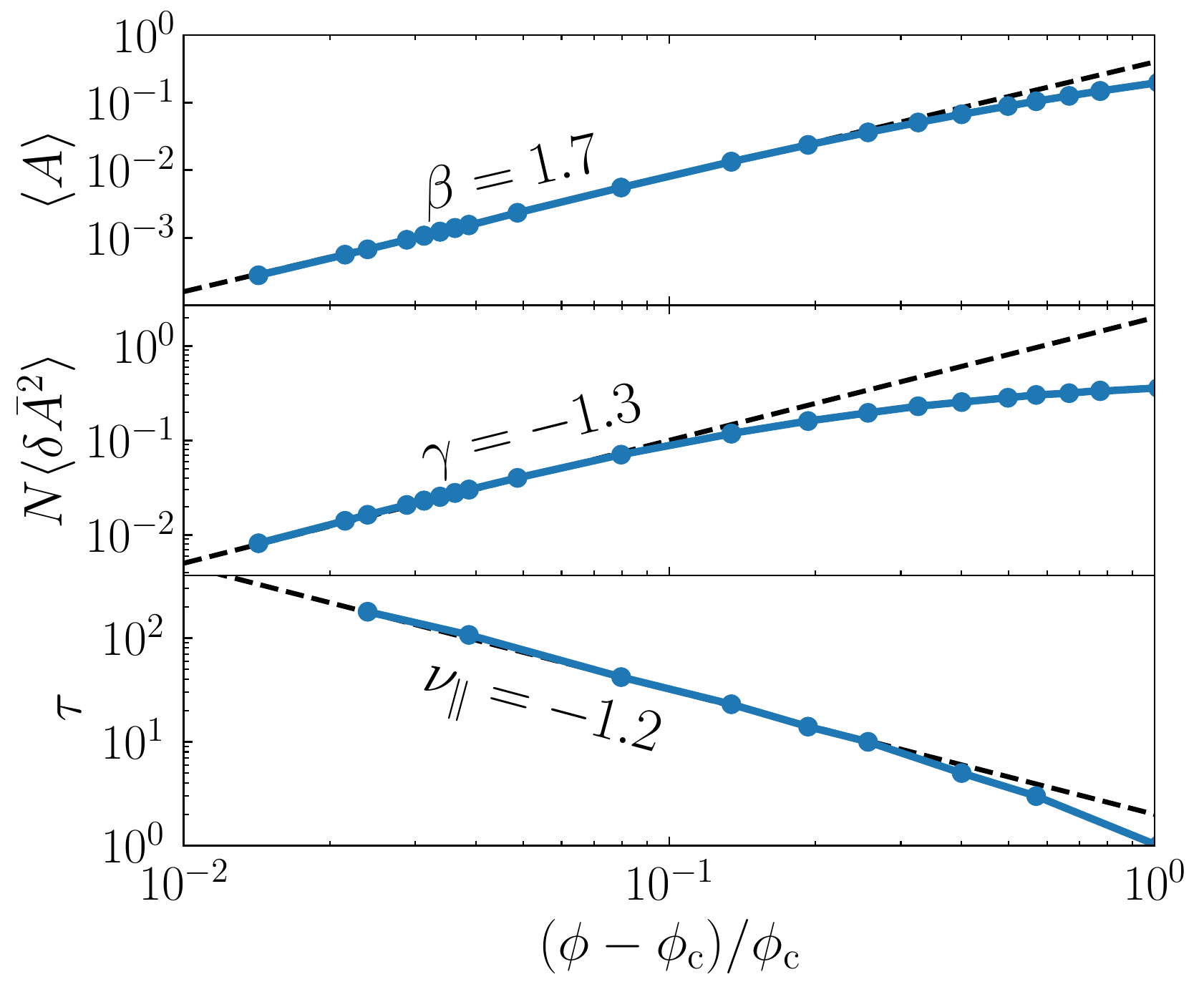}
\par\end{centering}
\caption{The average activity $\langle A \rangle$ (top), the normalized variance 
$N \langle (A-\langle A \rangle)^2\rangle$ of activity fluctuations (middle), and correlation time $\delta t$ of activity fluctuations (bottom),
versus $(\phi-\phi_c)/\phi_c$ for $c=0$ and $\lambda_\mathrm{p}/D = 1$.}
\label{fig:exponents_c0}
\end{figure}

We further characterize the new universality class in Fig.~\ref{fig:exponents_c0}, showing the critical behavior of the variance 
of the activity $N \langle \delta\bar{A}^2\rangle \sim \varepsilon^{-\gamma}$ where $\delta \bar{A} = \bar{A} - \langle A\rangle$
and of the activity correlation time $\tau \sim \varepsilon^{-\nu_\parallel}$, for $\lambda_\mathrm{p}/D = 1$. 
Here we define $\tau$ as $\langle \delta\bar{A}(t) \delta\bar{A}(t+\tau)\rangle/\langle \delta\bar{A}(t)^2 \rangle = 1/e$.
We find $\gamma \approx -1.3$ and $\nu_\parallel \approx -1.2$, in contrast to CDP values $\gamma_\mathrm{CDP} \approx 0.37$ and $\nu_{\parallel \mathrm{CDP}} \approx 1.3$.
The difference with CDP is such that even the sign of these exponents change; in our model, activity fluctuations vanish at the transition, while its correlation time diverges.
Both effects are a consequence of the convexity of the $\langle A \rangle(\phi)$ relation.

An important feature of the CDP class is the emergence of hyperuniformity, characterized by a scaling of the structure factor $S(\mathbf{q}) \sim q^{\kappa}$ when $q \to 0$ ($0<\kappa<1$) at $\phi_\mathrm{c}$~\cite{hexnerHyperuniformityCriticalAbsorbing2015,tjhung2015hyperuniform}: 
large scale density fluctuations are much weaker than for an equilibrium system at the same density. 
We show in~\cite{supp} that when increasing $\lambda_\mathrm{p}/D$, the hyperuniform regime observed in the TB model shrinks and a low-$q$ plateau develops, indicating that hyperuniformity is progressively smeared out by the diffusion induced by passive jumps.

\begin{figure}[t]
\begin{centering}
\includegraphics[width=1\columnwidth]{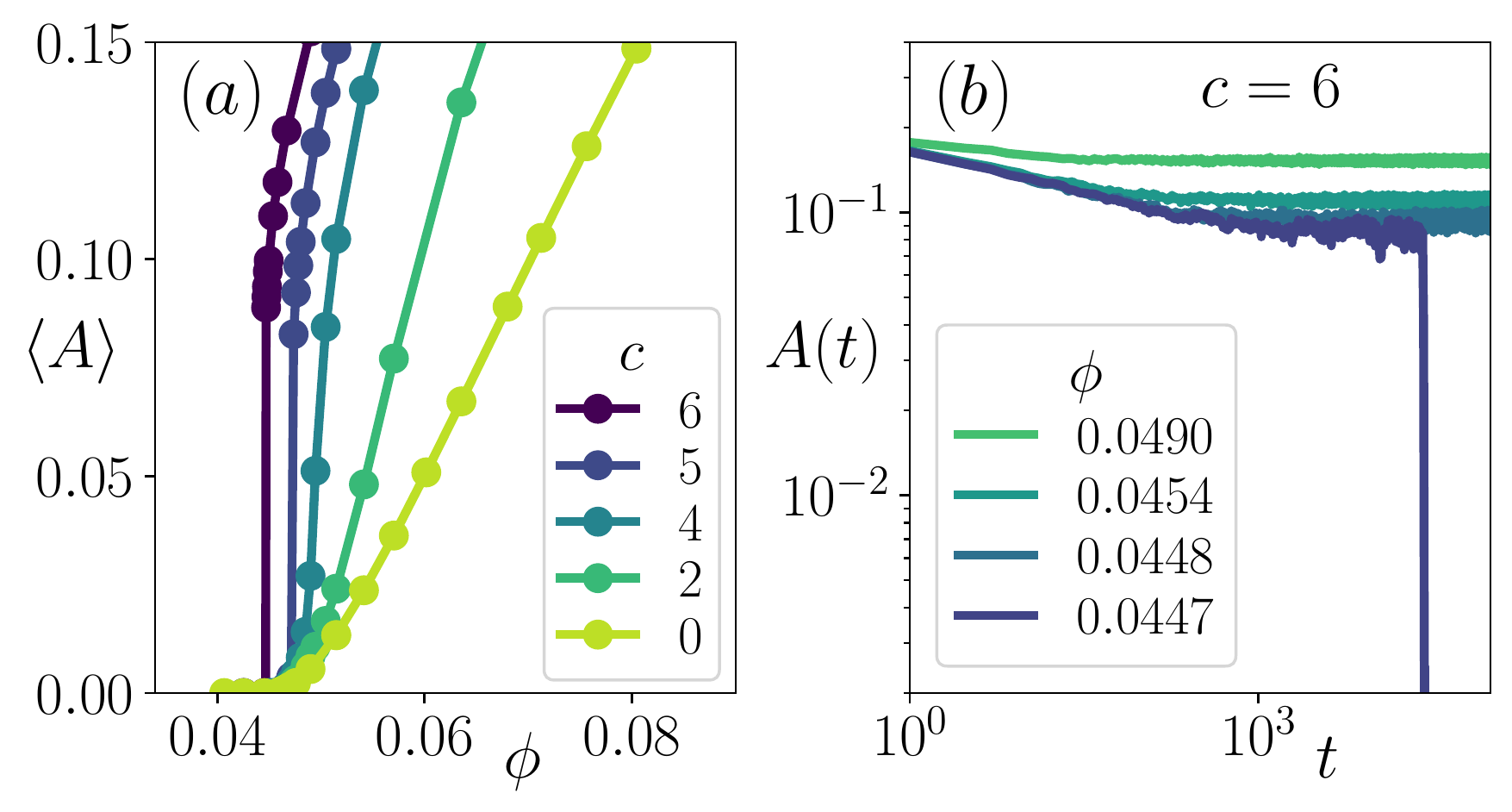}
\par\end{centering}
\caption{(a) Average activity $\langle A\rangle$ as a function of area fraction $\phi$ for $\lambda_\mathrm{p}/D = 1$ and several values of $c$. (b) Time series of the activity for the case $c=6$, for several $\phi$ across the first order transition.}
\label{fig:first_order}
\end{figure}

We now briefly comment on the case $c > 0$, shown in Fig.~\ref{fig:first_order}.
Increasing $c$ at fixed $\lambda_\mathrm{p}/D=1$, we observe that the curve $\langle A \rangle(\phi)$ becomes steeper close to the transition, eventually turning into a first-order transition [Fig.~\ref{fig:first_order}], where $\langle A\rangle$ jumps to a finite value at $\phi_\mathrm{c}$. 
Moreover, the relaxation to the absorbing state is discontinuous in time, with a first decay to a pseudo-steady plateau followed by a sudden collapse of the activity.
Due to finite size effects however, it is difficult to conclude whether the transition is a genuine APT, or if it is a discontinuous transition inside the active phase 
in the immediate vicinity of a continuous APT.
It is nonetheless reminiscent of observations in simulations of cyclically sheared particle model~\cite{kawasaki2016macroscopic,nagasawa_classification_2019}.

\paragraph{Continuum description.--}
The fact that passive jumps induce a new universality class for APTs with infinitely many absorbing states, distinct from CDP, is surprising. To rationalize this finding, we now look for a continuum description in a local mean-field framework. The crucial ingredient is the fact that the displacement of passive particles can create new active ones, 
and we thus expect the normal form $f(A)$ in (\ref{eq:Manna-standard-2}) to receive an additional contribution $f_\mathrm{p}(A)$ from passive particles. 
Estimating the number of active particles created in a time step $(t,t+\delta t)$ in terms of the radial distribution function of passive particles $g(r)$ leads to 
\begin{equation}\label{eq:mech}
N f_\mathrm{p}(A)\delta t 
\simeq
N_\mathrm{p} \rho_\mathrm{p}
\int d^d\bfr  \,
g(r) 
\mathcal{P}_{\rm overlap}(\Delta_\mathrm{p},r) \,.
\end{equation}
In the above expression, $r=|\bfr|$, $N_\mathrm{p}=N-N_\mathrm{a}$ and $\rho_\mathrm{p}=N_\mathrm{p}/N$ are the number of passive particles and their density, and  $\mathcal{P}_{\rm overlap}(\Delta_\mathrm{p},r) $ is the probability that a couple of particles at distance $r$ overlap in the next time step. Passing in polar coordinates and setting $r=D+\Delta_\mathrm{p} x$, we have
\begin{multline}\label{eq:mech-1}
\frac{\delta t f_p(A)}{2S_d\rho} = (1-A)^2 \Delta_\mathrm{p} \int_0^2 \mathrm{d}x [D+\Delta_\mathrm{p} x]^{d-1}  \\
 g (D+\Delta_\mathrm{p} x) \mathcal{P}_{O}(\Delta_\mathrm{p},x)\nonumber
\end{multline}
where $\mathcal{P}_{O}(\Delta_\mathrm{p}, x)=\mathcal{P}_{\rm overlap}(\Delta_\mathrm{p},D+\Delta_\mathrm{p} x) $, we used $\mathcal{P}_{\rm overlap}(\Delta_\mathrm{p},r>2(R+\Delta_\mathrm{p}))=0$, $S_d=d\pi^{d/2}/\Gamma(d/2+1)$ 
is the surface of the unitary sphere. We show in~\cite{supp} that for $d=2$  the expansion of  $\mathcal{P}_{O}(\Delta_\mathrm{p},x)$ in $\Delta_\mathrm{p}$ is 
\begin{equation}\label{eq:PO}
\mathcal{P}_{O}(\Delta_\mathrm{p},x) = \mathcal{P}_{O}^{(0)}(x) - \frac{\Delta_p}{D}  \mathcal{P}_{O}^{(1)}(x)+
\mathcal{O}(A),
\end{equation}
where $\mathcal{P}_{O}^{(i)}(x)$ with $i=1,2$ are given by integral expressions which can be readily evaluated numerically. It is important for what follows that $\mathcal{P}_{O}^{(i)}>0$ and that they do not depend on any parameter of the model.

The estimation of the radial distribution function $g (D+\Delta_\mathrm{p} x)$ for small $A$ is more subtle. To get insight, and assuming isotropy, we consider a minimal two-body description for the motion of two nearest neighbour passive particles $p_0$ and $p_1$. 
We fix $p_0$ at the origin and consider $p_1$ as a discrete-time random walker. The latter moves in an annular shape of radii $D$ and $L$ representing, respectively, $p_0$ and the second nearest particle to $p_1$. 
Whenever $p_1$ reaches one of the two boundaries, it is redistributed uniformly in the annulus. The pair correlation $g$ of the original model corresponds, in this effective description, to the stationary state $P_\mathrm{s}(r)$ of $p_1$. 
In three spatial dimensions, and for $L\to\infty$, $P_\mathrm{s}(D+\Delta_\mathrm{p} x) = \Delta_\mathrm{p}  Q_1(x) / (D +\Delta_\mathrm{p} x)$, where $Q_1(x)$ is given in terms of an inverse Laplace transform~\cite{supp} and, importantly, it does not contain any parameter of the model~\cite{majumdar2006unified,ziff2007general}. 
Surprisingly, the mathematical structure of the problem is very different in $d=2$ and, to the best of our knowledge, no exact solution is available. 
Yet, the numerical integration of this effective two-body problem is straightforward and its results are reported in~\cite{supp}. This allows us to conclude that, to a very good accuracy, $P_\mathrm{s}(D+\Delta_\mathrm{p} x) = \Delta_\mathrm{p} Q(x) / (D +\Delta_\mathrm{p} x)$ also for $d=2$, and $Q(x)$ is independent of $\Delta_\mathrm{p}$ and $c$. We thus conclude that 
\begin{equation}\label{eq:g_Delta}
(D+\Delta_\mathrm{p} x)g (D+\Delta_\mathrm{p} x)\sim_{\Delta_\mathrm{p}\to0} \Delta_\mathrm{p}\,\,g^{(0)}(x) \,
\end{equation}
where, obviously, $g^{(0)}(x)>0$ for all $x$.
Combining (\ref{eq:mech}) and (\ref{eq:g_Delta}) we conclude that
\begin{equation}\label{eq:mech-2}
f_\mathrm{p}(A)  = \alpha_\mathrm{p} \rho A - \mu \rho A^{3/2} + \mathcal{O}(A^2)\,
\end{equation}
where $\alpha_\mathrm{p}=(2S_d/\delta t)\lambda_\mathrm{p} 
\int_0^2 \mathrm{d}x \,g^{(0)} \mathcal{P}_{O}^{(0)}>0$ 
and $\mu=(2S_d/\delta t)(\lambda_\mathrm{p}^2/D)\int_0^2 \mathrm{d}x \,g^{(0)} \mathcal{P}_{O}^{(1)}>0$; 
the normal form in (\ref{eq:Manna-standard-2}) should thus be 
\begin{equation}\label{eq:new-normal-form}
f(A) = (-\alpha+\tilde{k}\rho)A -\mu\rho A^{3/2} +\mathcal{O}(A^{2})\,,
\end{equation}
where $\tilde{k}=k+\alpha_p$.
The leading term of (\ref{eq:new-normal-form}) is merely a renormalization of the linear coefficient of (\ref{eq:Manna-standard-2}) and it does not change the universal critical properties of the APT.
Mediated interactions however have a more drastic effect, as the normal form acquires a $A^{3/2}$ contribution. Because $\mu>0$, we predict that the APT remains continuous at mean-field level, 
but with an order parameter exponent $\beta_\mathrm{MF}=2$, a value slightly larger than the numerically measured value $\beta \approx 1.7$. 

The coarse-graining presented above strongly supports the existence of a new universality class for APT whenever infinitely many absorbing states are present and local activity affects passive particles. It further provides the field-theoretical description within which such new universality class might be studied. This corresponds to replacing the CDP normal form (\ref{eq:Manna-standard-2}) with (\ref{eq:new-normal-form}). It should be also noted that the density evolution (\ref{eq:Manna-standard-1}) is expected to be transformed into 
\begin{equation}
\p_t \rho = D_{\rho}\nabla^2 A
+D_m  \nabla^2(A \rho) +\sigma_m\nabla\cdot\left(\sqrt{A\rho}\,\bm{\xi}\right)
\label{eq:Manna-modified-1}
\end{equation}
with $\bm{\xi}$ a vectorial Gaussian white noise. While dimensional analysis indicates that $D_m$ and $\sigma_m$ are irrelevant close to the upper critical dimension, a detailed renormalization group analysis or large scale numerics of our continuum theory are needed to assess the importance of $D_m$ and $\sigma_m$; this is left for future works.   
We finally observe that the presence of a discontinuous APT numerically found at high $c$ values is likely explained by a change in the sign of $\mu$ (still assuming a stabilising $A^2$ term). In fact, direct measurements of $g^{(0)}$ reported in~\cite{supp} indicates that, in the many body system, $g^{(0)}$ is not independent of $A$ exactly: it can be shown that this adds a new contribution to $\mu$ which might change its sign.

\paragraph{Conclusion.--}
We investigated a minimal model for the stroboscopic dynamics of periodically sheared soft matter, taking into account the effect of active regions on passive ones in a mean-field way. 
We found that the presence of these mediated interactions modifies the universality class of the APT, which does not belong to CDP anymore. 
We characterized the APT using extensive numerical simulations and a local mean-field analytical argument to derive a continuum theory. 
Our results show that CDP is not the only universality class of APTs with infinitely many absorbing states; as such, the field-theoretical description proposed here is expected to describe APTs in many other contexts in which local activity affects nearby passive regions. 

Here, the displacement of passive particles is assumed to depend on the spatially averaged activity, in a mean-field spirit. In a realistic system with mediated interactions, this  assumption seems relevant when the propagator of the interaction is long-ranged. For short-range propagators (e.g.~screened hydrodynamic interactions in dense suspensions), displacements of passive particles rather result from a local activity.
Whether this change would modify the universality class of the absorbing phase transition is an important question left for future work. 

In any case, our results indicate that mediated interactions present in experiments and realistic numerical models are a crucial physical ingredient that make the APT depart from the CDP class. 
In practice, it is often difficult  to distinguish active particles from passive ones because all particles move, even ever so slightly~\cite{jeanneret_geometrically_2014,himanagamanasaExperimentalSignaturesNonequilibrium2014}. 
This could be a consequence of mediated interactions. A test of this could be to determine the distribution of particle displacements. Close to the APT, this distribution is expected to become bimodal, with much smaller displacements for passive particles than for active ones.
Finally, our results imply that hyperuniformity is suppressed by mediated interactions, a consequence of experimental relevance since scattering techniques may provide access to the structure factor in experimental systems.

{\em Acknowledgements.} We thank M.E. Cates, C. Ness, E. Tjhung for discussions. 
CN acknowledges the support of an Aide Investissements d'Avenir du LabEx PALM (ANR-10-LABX-0039-PALM).


%


\end{document}